\documentclass[12pt]{article}

\usepackage{amsmath,amssymb,graphicx}
\usepackage[final]{hyperref}
\usepackage{mathrsfs}
\usepackage{dcolumn}
\usepackage{bm}
\usepackage{cite}
\usepackage{times}

\hypersetup{
colorlinks=true, 
linkcolor=blue, 
citecolor=blue, 
filecolor=magenta, 
urlcolor=blue
}
\newenvironment{sciabstract}{%
\begin{quote} \bf}
{\end{quote}}

\title{Wave-Front Reconstruction via Single-Pixel Homodyne Imaging}

\author{Savannah L. Cuozzo,$^{1}$ Charris Gabaldon,$^{1}$ Pratik J. Barge,$^{2}$
Ziqi Niu,$^{1}$ \\Hwang Lee,$^{2}$ Lior Cohen,$^{2}$ Irina Novikova,$^{1}$ and Eugeniy E. Mikhailov$^{1,*}$\\
\\
\small{$^{1}$William \& Mary, Department of Physics, 300 Ukrop Way,
Williamsburg, VA 23185}\\ 
\small{$^{2}$Louisiana State University, Department of Physics, 
202 Nicholson Hall, Baton Rouge, LA 70803} \\
\small{$^{*}$eemikh@wm.edu}
}

\date{}

\begin{document}
\baselineskip24pt

\maketitle

\begin{sciabstract} We combine single-pixel imaging and homodyne detection to perform full object recovery (phase and amplitude). Our method does not require any prior information about the object or the illuminating
fields. As a demonstration, we reconstruct the optical properties of several
semi-transparent objects  and find that the reconstructed complex transmission has a phase precision of 0.02 radians and a relative amplitude precision of 0.01. 
\end{sciabstract}

\section{Introduction}
Full object recovery (i.e. amplitude and phase) is extremely
useful in multiple applications. The intrinsic limitation of light
detectors -- sensitivity to only incoming light energy -- makes it
impossible to infer both amplitude and phase with only one measurement.
Several numerical methods were developed to infer the full
wavefront~\cite{russell2002siamWavefrontReconstructionReview}, they
generally require several measurements via scanning the distance and
observing the modified intensity profile and then coupling it with iterative
approximation routine for the reconstructed wavefront. Such methods are
usually computationally intensive~\cite{Wang2019} and often requires some
{\it a priori} information about the wavefront modifying object. For
example, holography~\cite{Wittwer2022,
Clemente_Single_Pixel_Holography_2013, Indebetouw2006} and
ptychography~\cite{Li2021, Wittwer2022, Sidorenko2016} assume only small
modification of wavefront amplitude.

Reconstruction of the spatial object complex transmission coefficient, i.e.
wavefront modification right behind the object requires either object
raster or collecting speckle patterns with images. This makes it time
consuming, and thus sensitive to illumination instabilities, or increase
data storage demand to collect multiple images of the speckle pattern. Use
of cameras makes the reconstruction method suitable only for a subset of
wavelengths, dimming the use of the speckle pattern method at X-ray or THz
wavelengths. These problems are circumvented by the use of single-pixel imaging
(SPI) techniques~\cite{Gibson2020} where the object illuminated (or sampled
at the detector) with multiple spatial light profiles allowing to reconstruct
its 2D shape~\cite{Li2021,Lee2010}.  Expansion of SPI ideas to the phase
shifting holography~\cite{Yamaguchi97ol_digitalHolography} allows to
reconstruct full wavefront with the assumption that a reference beam is
spatially uniform in amplitude and
phase~\cite{Clemente_Single_Pixel_Holography_2013,
	Santos2021ao_single_pixel_comparison,
	Hou2021oe_complex_amplitude_spi}. The SPI method alongside with the
	compressive sampling techniques~\cite{Tao2006IEEE,
	2008BaraniukIEEE, Donoho2006IEEE, Lee2010,
Hou2021oe_complex_amplitude_spi} can significantly reduce the required
the number of measurements.

We present an alternative method for the reconstruction of an object's spatial
complex transmission (amplitude and phase) that relies on using
phase-sensitive amplification from homodyne detection and the spatial
information reconstruction of single-pixel imaging. Unlike previously
reported scanning homodyning~\cite{Eang2015} our method does not have physical
motion of either sample or probing beams. Our treatment of the system is
more general and does not rely on the commonly used assumption of the spatial
uniformity~\cite{Eang2015,Hou2021oe_complex_amplitude_spi,Santos2021ao_single_pixel_comparison}
of reference or local oscillator (LO) beams. An additional benefit of our
method is the ability to work with extremely small probe light intensity
(theoretically down to a single photon level) owing to excellent LO noise
suppression and amplification by homodyning. Such ability to image with
a weak probe is extremely useful for bioimaging which is sensitive to the
photo-induced damage of biosamples.

\section{Theory of single-pixel imaging via homodyne detection} 

In traditional single-pixel imaging (SPI), a light source illuminates a
scene and is then collected onto a photodiode using a form of structured
detection~\cite{Gibson2020}. Structured detection involves projecting your
sampled scene onto a set of spatial basis elements, $H_i$, and then
measuring the intensity transmitted with a photodiode (single-pixel
camera), $w_i$. The scene can then be reconstructed, pixel by pixel, based
on the intensity measured for each basis element, $\mathrm{Image} =
\Sigma_i w_i H_i$. SPI works well to reconstruct the image intensity.
However, if we desire to reconstruct the full wavefront (i.e. phase
information), we will need to include another layer of detection that is
sensitive to phase. The digital holography uses a 4-step phase shift
technique, to obtain phase-dependent (complex) weights required for the
full wavefront reconstruction~\cite{Lee2010,
Yamaguchi97ol_digitalHolography, Hou2021oe_complex_amplitude_spi}, but it
works only in the assumption of the spatially uniform reference beam.
We have chosen to combine SPI with
homodyne detection -- single-pixel homodyne imaging -- to circumvent such
limitations.

Phase-sensitive \textit{homodyne detection} allows us to probe optical depth while also amplifying a weak probe beam. Homodyning requires another beam as a phase reference, 
traditionally called the local oscillator (LO). The LO and interrogating probe (which illuminates our sample) are mixed on a 50/50 beam splitter and the resulting fields are measured with balanced photodiodes. The two beams are required to be phase-locked.
For our imaging method, it is important to consider a general EM field shape of the LO ($u_{LO}(x,y)e^{i\Delta\phi}$) and interrogating probe field ($u_{in}(x,y)$), where $x$ and $y$ are the spatial positions, $u$ is the complex amplitude of the field, and $\Delta\phi$ is the relative phase between the LO and probe.

Using the standard homodyning formalism \cite{bachor_guide_2004}, with the assumption $|u_{LO}|>>|u_{in}|$, we can write the differential current from the balanced photodiode, $i_{d}(\Delta\phi)$, as
\begin{equation}
\label{eq:differentialcurrent}
		i_{d}(\Delta\phi)\sim \int_A\left( u_{LO}u^{*}_{in}+u^{*}_{LO}u_{in}\right)ds
		=2Re(\mathcal{O}(\Delta\phi)),
\end{equation}
where we define the overlap, $\mathcal{O}$, as
\begin{subequations}
\begin{align}
\mathcal{O} &= \int_A u^{*}_\mathrm{LO}u_\mathrm{in}ds,\\
\Delta\phi &= \phi_{in} - \phi_{LO} = \arg(\mathcal{O}).
\end{align}
\label{eq:overlap}
\end{subequations}
Here $\phi_{in}$ and $\phi_{LO}$ are phases of the interrogating probe and
LO fields.
In the homodyne detection scheme, the probe field
is amplified by the LO, only contributing to the final signal when the LO and input field are spatially overlapped.

The balance photodiode zeroes the large DC component proportional to the LO
or reference intensity which is present in holography
method~\cite{Yamaguchi97ol_digitalHolography,Clemente_Single_Pixel_Holography_2013}.
Thus, homodyning eases the requirement of the high dynamic range of the
detector acquisition, since the interfering term of
eq.~\ref{eq:differentialcurrent} can be quite small in comparison to DC
term.
Combined with homodyning, we are also insensitive to the noise of the LO
while amplifying the weak input field $u_{in}$.

By utilizing the phase-dependent nature of homodyne detection in conjunction with \textit{single-pixel imaging}, we can
reconstruct the field and move beyond simple intensity reconstruction.
Single-pixel imaging alone relies on sampling an image with a known
structured basis and then reconstructing based on the intensities measured
at a single photodiode. In the same way, our method relies on sampling the
object of interest by using structured detection, i.e. we shape the LO in
the set of the orthogonal modes and
amplify portions of our probe beam based on the overlap with the LO.

\begin{figure} 
	\centering 
	\includegraphics[width=\linewidth]
	{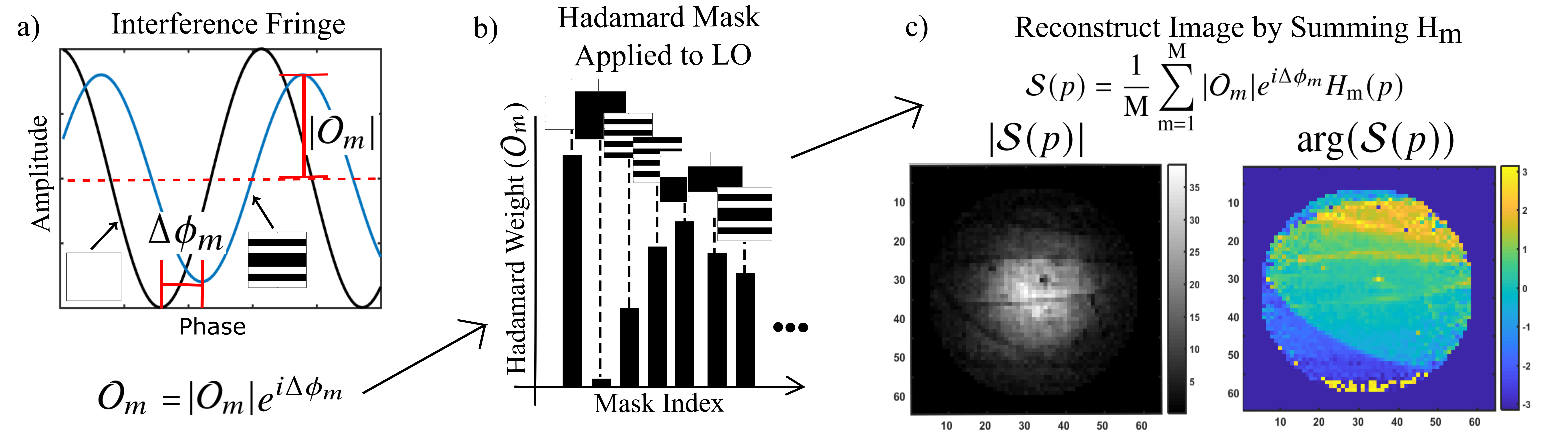}
	\caption{A conceptual visualization of the image reconstruction process.
		a) Extraction of the phase ($\phi_m$) and amplitude
		($|\mathcal{O}_m|$) of the overlap parameter from
the LO phase-dependent differential current of the homodyne detector. For
every LO mask, the phase is calculated with respect to the same reference
mask (we used the ``all-pixel-on'' mask, i.e a mask of all 1s).
b) For a given Hadamard mask ($H_m$), we extract a complex weight
($\mathcal{O}_m$).
c) Combining weighted Hadamard masks, we reconstruct the sampled field
product, $S(p) = u_{LO}^*(p) u_{in}(p)$.
These colored maps show the field product reconstruction in the situation
when both LO and probe beams
are in the fundamental Gaussian mode but the probe beam passed through an
insect wing. Different phases in the reconstruction indicate the variable
thickness of the wing.} 
	\label{fig:processingflow}
\end{figure}

It is possible to reconstruct with any complete sampling basis. We choose to use the Hadamard matrices as our basis, which are 
formed with a $\pm1$ tiling where the rows of each matrix are orthogonal to 
the others. 
The differential homodyne signal (Eq.~\ref{eq:differentialcurrent})
is governed by the overlap with each of the $m$ basis modes of the LO (see
Fig.~\ref{fig:processingflow} a and b).
The overlap in the real $xy$ space can be approximated in the pixels space 
as the sum, $\mathcal{O} = \sum_m \mathcal{O}_m$, where 
\begin{equation}
	\mathcal{O}_m = \sum_p u^*_{LO}(p) u_{in}(p) H_m(p)
	\label{eq:discreteoverlap}
\end{equation}
where $\mathcal{O}_m$ is the overlap of the LO, probe, and $m$th Hadamard mode,
and $p$ goes over the pixels of the Hadamard mask. When the masked LO has a strong overlap with the
input field, the differential signal will increase.
In the experiment, we find $\mathcal{O}_m$ by sweeping through all possible LO phases and
fitting the observed differential current to modified 
Eq.~\ref{eq:differentialcurrent} which is now LO mode specific:
\begin{equation}
\label{eq:ReO}
i_{d,m}(\Delta\phi_m) = i_{d,m}(\phi_m - \phi_{LO}) \sim 2|\mathcal{O}_m|\cos(\Delta\phi_m).
\end{equation}
We can extract both the overlap, $|\mathcal{O}_{m}|$, and phase,
$\Delta\phi_{m}$, which allows full reconstruction of the wavefront
product, $\mathcal{S} = u^*_{LO}u_{in}$, by the following
formula, similar to intensity-based single-pixel imaging:
\begin{equation} 
\mathcal{S}(p) = \frac{1}{\mathrm{M}}
\sum_{\mathrm{m} = 1}^\mathrm{M} |\mathcal{O}_{m}|e^{i\Delta\phi_m}H_{\mathrm{m}}(p).
\label{eq:classicalrecon}
 \end{equation} 
where $p$ is the pixel of the Hadamard mask, $H_m$ is the Hadamard mode,  $\mathcal{O}_m$ is the overlap between the LO and probe field,
and $\Delta\phi_m$ is the relative phase of the shaped LO and probe field. See Fig.~\ref{fig:processingflow} for a visual depiction of the image reconstruction process. It is worth mentioning that, our method is compatible with compressive sensing techniques, which allow fewer samples than required for completeness. We performed numerical simulations to confirm this. 

We define object {\em complex} transmission, $T$, as the ratio of the reconstructed wavefront, $\mathcal{S}_{\mathrm{obj}}(p)$, that has probed an object and a reference, $\mathcal{S}_{\mathrm{ref}}(p)$, which did not,
\begin{equation}
T(p) = \frac{\mathcal{S}_{\mathrm{obj}}(p)}{\mathcal{S}_{\mathrm{ref}}(p)}.
\label{eq:transmission}
\end{equation}

\section{Experimental Methods}
\begin{figure}
    \centering
    \includegraphics[width =\linewidth]{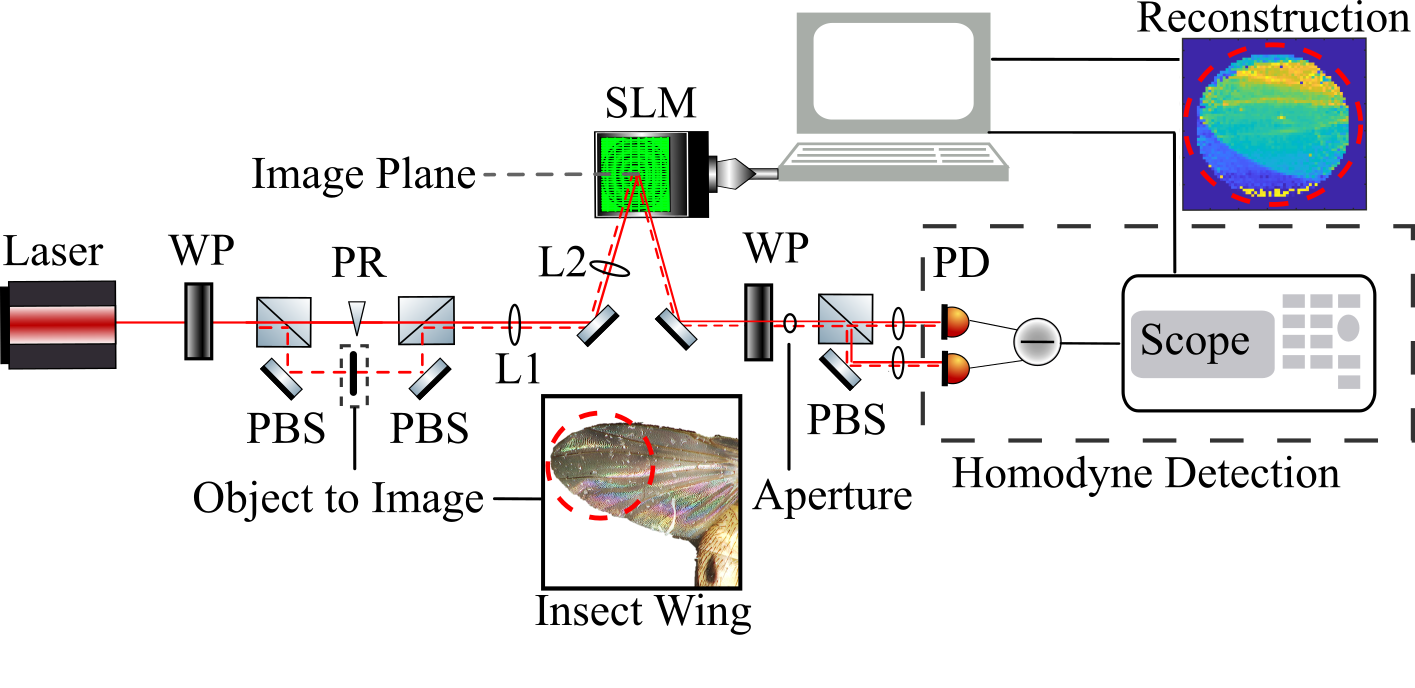}
    \caption{A schematic of the experimental setup where, WP is a $\lambda/2$ wave plate, PBS is a polarizing beam splitter, PR is a phase retarder, L1 and L2 are imaging lenses (L1 = 100 mm, L2 = 300 mm), SLM is a spatial light modulator, and PD is the balance photodiode. In this example, we image an insect wing and show the reconstructed phase.}
    \label{fig:expsetup}
\end{figure}

A schematic of our experimental setup can be found in
Fig.~\ref{fig:expsetup}. The beam used in our setup is produced by a 795 nm
laser.  We spatially filter the beam to have a predominately Gaussian
spatial profile and then use a beam splitter to separate a small amount of
light which acts as our interrogating probe beam, $u_{in}(x,y)$.
The rest of the beam is utilized as the LO, $u_{LO}(x,y)e^{i\phi_{LO}}$.
The probe samples an object and is then spatially overlapped with the LO
(Eq.~\ref{eq:overlap}a).

The two beams travel collinearly through a 4-f imaging system, which moves
the imaging plane of the object onto a spatial light modulator (SLM) plane.
Although both beams reflect off the SLM, only the LO's spatial profile is
modified, because the SLM is polarization sensitive and the beams are in
orthogonal polarizations. Having both beams travel together through
most of the optical path allows us to maintain the relative phase stability
of the probe and LO. 

To modify the overlap between the LO and probe, and extract information
about our object according to Eq.~\ref{eq:discreteoverlap}, we place a
series of Hadamard masks ($H_{m}$) onto our LO using the SLM (Meadowlark
Optics SLM with 512x512 pixels). This SLM only modifies the phase of our
beam, but by applying a blazed grating~\cite{Boyd2013ol_slm_profiling}
to a group of least 5x5 pixels, we were able to generate required
intensity mask with effective 64x64 pixels.
The Hadamard matrices, which we use as the detection basis, consist of $\pm1$
tiling, the SLM cannot physically generate a negative intensity pixel. To work around this
technical hurdle, we break our true Hadamard mask into a linear combination
of two masks containing only 1s and 0s, and then subtract the weighted
pairs similar to \cite{Gibson2020}. 

We controllably change the path length (phase) between the LO and the probe by driving a piezoelectric (PZT) mounted mirror in the interferometric path. This effectively acts as a phase retarder (PR). We sweep the phase between the LO and probe and record the interference fringe for each of the Hadamard masks applied by the SLM. 

After a mask is applied to the LO the two beams are mixed on a
beamsplitter and propagate to the far field where they are collected onto
two balanced photodiodes.  The differential current
(Eq.~\ref{eq:differentialcurrent}) allows us to extract the phase of each
Hadamard mask. By fitting the fringe, we can extract the amplitude and
relative phase information for a specific mask (see Eq.~\ref{eq:ReO} and
Fig.~\ref{fig:processingflow}a) which is used to construct the weight
defined in Eq.~\ref{eq:classicalrecon}. We use a reference phase to correct
for any slow interferometric drifts.

\section{Results}
\begin{figure} 
	\centering 
	\includegraphics[width=\linewidth]
	{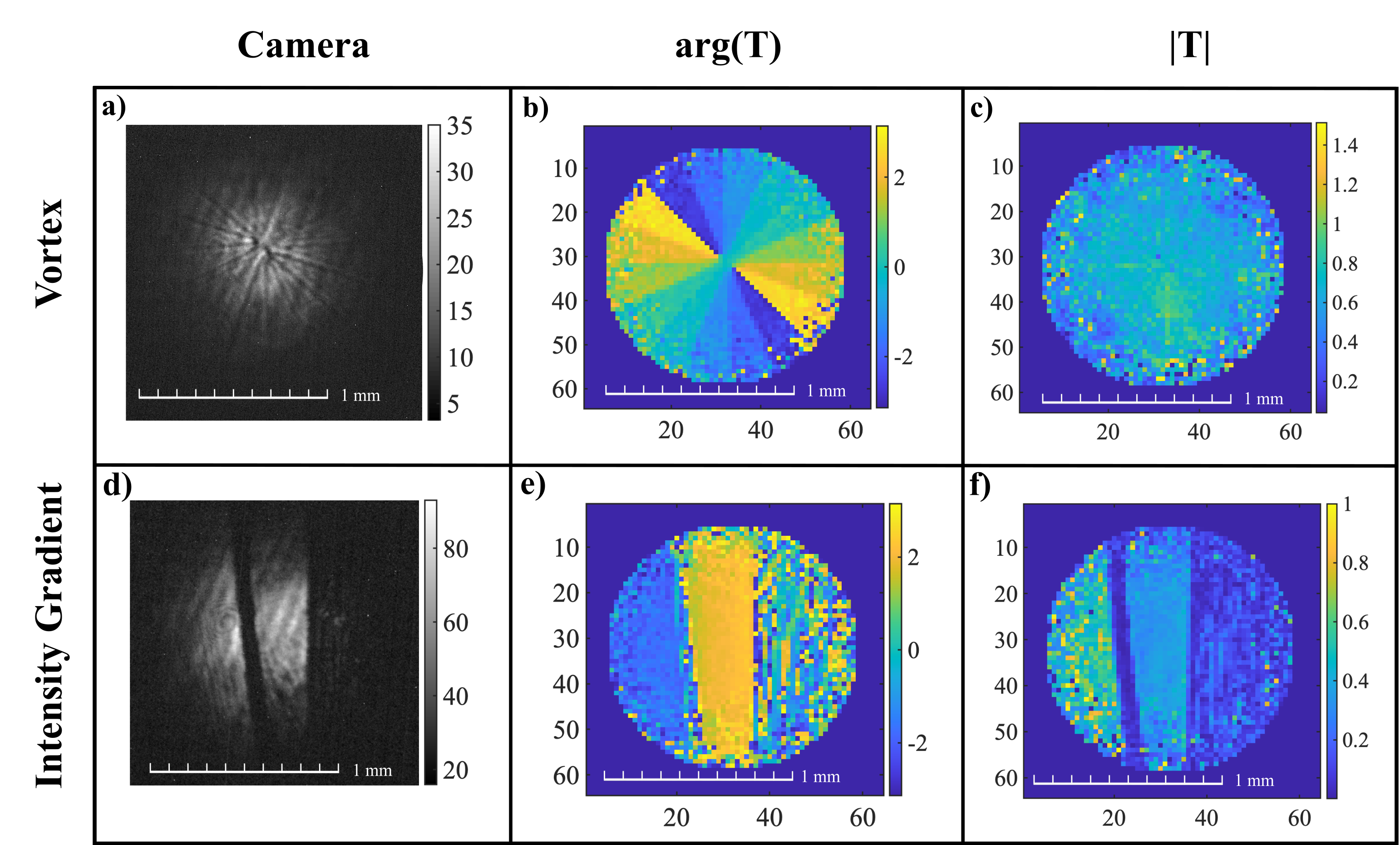} 
	\caption{We compare traditional camera images of a) an optical
		vortex mask with strong illumination and d) an intensity gradient
		where the left-hand side contains no object, the center has a uniform
		absorptive filter and the right-hand side contains two absorptive filters.
		b) The phase reconstruction of the vortex mask resolves various depths that
		are otherwise minimal in the c) amplitude reconstruction. e) The phase
		reconstruction of the intensity gradient object shows a defined phase in
		the center and the left-hand side that correspond to the ND filter and
		empty beam, respectively. The right-hand side with multiple absorptive
		filters is undefined due to the small amount of light transmitted. f) The
		amplitude reconstruction of the intensity gradient shows more light is
		transmitted through the left, empty area compared to the center with one
		filter. The three object areas are differentiable and we can see the amount
		of light transmitting through decreases as we move across the beam from
	left to right. }
	\label{fig:summary}
\end{figure}

\begin{figure} 
	\centering 
	\includegraphics[width=\linewidth]
		{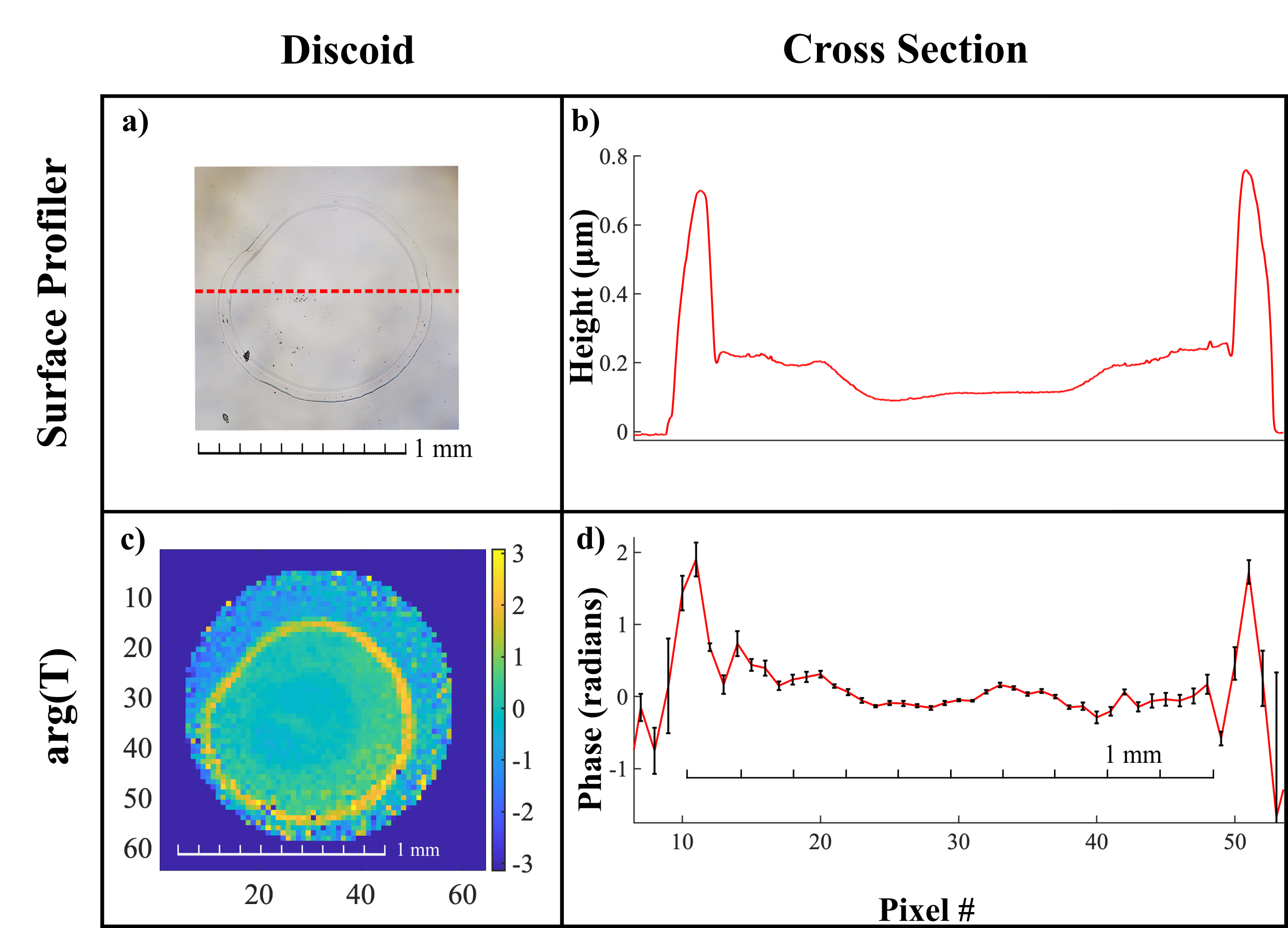} 
	\caption{a) A HIROX microscope system image of the discoid, after the alcohol evaporated, taken in mid-range at x140 (field of view is 2169.05$\mu m$ with a resolution of 1.13 $\mu m$). The horizontal dashed red line indicates the approximate b) cross-section taken with a Bruker Dektak Surface Profiler. The vertical uncertainty is 0.1 nm. c) The phase profile of the discoid outlines the ridge and crater-like shape that is physically present. d) A similar cross-section of our phase reconstruction is qualitatively compared to the Bruker Dektak Profiler cross-section. The edges have a larger uncertainty, which is attributed to the minimal light at the edges of the beam as the signal-to-noise ratio decreases.}
	\label{fig:crosssection}
\end{figure}

To demonstrate the phase sensitivity of our method, we choose to reconstruct several transparent objects with variable optical depths (see Fig.~\ref{fig:summary}). The background of each object reconstruction is removed by dividing out an empty beam from the object, resulting in a phase and amplitude transmission masks (Eq.~\ref{eq:transmission}).

The top row of Fig.~\ref{fig:summary} shows experimentally reconstructed images of an optical vortex mask which is a glass plate with sector-wise variable thickness spanning 850nm (Fig.~\ref{fig:summary}(b,c) ).
In our phase transmission reconstruction (Fig.~\ref{fig:summary}b), we observe clear phase steps corresponding to the sectors of the phase plate. Notice that no such information is available in the direct camera image (Fig.~\ref{fig:summary}a), which is sensitive only to changes in intensity. As expected, the amplitude transmission reconstruction shows a flat profile with an approximate transmission of 1 (Fig.~\ref{fig:summary}c). Due to the noise around the edges of the beam, the maximum transmission is scaled larger than 1.

Fig.~\ref{fig:summary}d shows a traditional camera image of the intensity
gradient mask where the left-hand side is empty, the center contains one
absorptive filter, and the right-hand side contains two overlaid absorptive
filters. The empty left side and the center filter object appear to be very
similar to the traditional camera image (Fig.~\ref{fig:summary}d ) but the
reconstructed amplitude transmission  (Fig.~\ref{fig:summary}f) shows less
light is transmitted through the center.  Our reconstructed phase transmission profile (Fig.~\ref{fig:summary}e) clearly shows a distinction between the left and center sections but is undefined on the right side due to low light as the light transmission is small.

To qualitatively analyze the depth resolution of our phase transmission
reconstructions, we take several data sets of the same object, a drop of
sanitizer which after the alcohol evaporation has a discoid shape. This is
compared to a traditional surface measurement method
(Fig.~\ref{fig:crosssection}). Using a Bruker Dektak Surface Profiler, a
diamond-tip stylus sweeps across the surface of the discoid to record the
height differences between the ridge circumference boundaries, showing a
crater-like depth shape. Fig.~\ref{fig:crosssection}b shows the cross
section result of this method with a vertical resolution of 0.1 nm. The
Bruker Profiler performs a standard scan over a 2 mm length for a total of
20 seconds. The stylus has a 2 $\mu m$ radius and uses a force equivalent
to 3 mg. The
phase transmission (Fig.~\ref{fig:crosssection}c) of the discoid is
reconstructed multiple times and a similar cross-section
(Fig.~\ref{fig:crosssection}d) is taken across the crater-like shape. The
uncertainty of the phase transmission reconstruction comes from the
statistics of the multiple runs.

Although slightly different cross-sections are taken of the microscope image and the reconstructed phase, there is a general agreement of the discoid shape and the perimeter locations. 

Based on multiple data sets of discoid phase reconstruction
(Fig.~\ref{fig:summary}c) the experimental phase reconstruction uncertainty
is $\pm 0.02$  radians in the center of the image where we have the strong LO.
We can reconstruct the amplitude of transmission with a statistical uncertainty of $\pm0.01$.
However, the phase and amplitude sensitivity of our method is ultimately
limited by the shot noise level, though in our case we were limited by
the phase stability, quantization noise of our analog to digital
converter,  and the dark noise of our detector.
Spatially we can reconstruct images with
64x64 pixels with the physical size of the pixel corresponding to $26 \pm 2 \mu m$.
But, no physical limitations are keeping our method
from approaching the Rayleigh limit, in terms of spatial resolution.

\section{Conclusion} 

We demonstrate how the use of single-pixel imaging with homodyne detection
can expose full transmission information (amplitude and phase) about the
object. This method is simple in design yet requires no assumptions about
the objects' characteristics or illuminating light before imaging. The ability to
use very weak probe light makes our method attractive for bioimaging. Additionally, it is worth mentioning that, our method is compatible with compressive sensing techniques.   

\section{Funding}
Air Force Office of Scientific Research (FA9550-19-1-0066).

\section{Acknowledgments}
We thank Olga Trofimova and Doug Beringer from William \& Mary Applied
Research Center (ARC) Core Labs for assisting us with the characterization  of
our samples with Bruker Dektak XT Surface Profiler  and HIROX RH-2000 High Resolution Digital-Video Microscopy System.


\bibliography{bibliography}
\bibliographystyle{Science}
\end{document}